\documentstyle[12pt,a4]{article}

\begin{document}

\parindent=16pt

\newtheorem{theorem}{Theorem}
\newtheorem{lemma}{Lemma}
\newtheorem{defn}{Definition}

\newcommand{\be}{\begin{eqnarray}}
\newcommand{\ee}{\end{eqnarray}}
\newcommand{\bes}{\begin{eqnarray*}}
\newcommand{\ees}{\end{eqnarray*}}
\newcommand{\beqn}{\begin{equation}}
\newcommand{\eeqn}{\end{equation}}

\newcommand\Pa{Painlev\'e}

\title{The Discrete Painlev\'e I Hierarchy}
\pagestyle{empty}
\author{Clio Cresswell\\
         {\it  School of Mathematics}\\
         {\it University of New South Wales}\\
         {\it  Sydney NSW 2052, Australia}\\
         and\\
         Nalini Joshi\\
         {\it Department of Pure Mathematics}\\
         {\it University of Adelaide}\\
         {\it  Adelaide SA 5005, Australia}
}\date{}
\maketitle

\begin{center}
{\bf Abstract}
\end{center}

The discrete Painlev\'e I equation (dP$\rm_I$)
is an integrable difference equation which has the classical
first Painlev\'e equation (P$\rm_I$) as a continuum limit. dP$\rm_I$
is believed to be integrable because it is the
discrete isomonodromy condition for an associated (single-valued)
linear problem.
In this paper, we derive higher-order difference equations
as isomonodromy conditions that are
associated to the same linear deformation problem.
These form a hierarchy that may be compared to hierarchies of integrable
ordinary differential equations (ODEs).
We strengthen this comparison by continuum limit calculations
that lead to equations in the P$\rm_I$ hierarchy.
We propose that our difference equations are
discrete versions of higher-order Painlev\'e equations.

\section[1]{Introduction}

Our aim is to derive higher-order versions of the equation
\be
        y_{n+1}+y_n+y_{n-1}={\alpha n+\beta\over y_n} +\gamma .
        \label{dPI}
\ee
Equation (\ref{dPI}) is known as the first discrete Painlev\'e equation or
dP$\rm_I$,
because the scaled continuum limit
$y_n=1+h^2 u(x)$, $x=nh$, with
$\alpha=r_1h^5$, $\beta=-3+r_2h^2$,
$\gamma=3-\beta-r_3h^4$, where the $r_i$ ($i=1,2,3$)
 are constants, yields a scaled and translated version of the classical
first Painlev\'e equation (P$\rm_I$)
\[u''=6u^2+x,\] as $h\to 0$.

P$\rm_I$ is the simplest of six  well known nonlinear
second-order ODEs
in the complex plane called the Painlev\'e equations.
Their characteristic property that all
movable singularities of all solutions are poles is called the Painlev\'e
property.
Painlev\'e \cite{pp:acta}, Gambier \cite{gambier:acta}, and
Fuchs \cite{fuchs:annalen}
identified them (under some mild conditions) as the only
such equations with the Painlev\'e property whose general solutions
are new transcendental functions.

P$\rm_I$ is believed to be integrable because it is the isomonodromy
condition for an associated (single-valued)
linear system of differential equations
\cite{okamoto:86}. dP$\rm_I$ is also an isomonodromy condition.
Moreover, it possesses a discrete version of the Painlev\'e property
called the singularity confinement property proposed by Grammaticos,
Ramani \textit{et al}\cite{Gram:Ram:Papa,Ram:Gram:Hiet}.

Joshi \textit{et al} \cite{Josh:iso} derived the linear problem associated
with dP$\rm_I$
by starting with the Ablowitz-Ladik \cite{Ab:La,Ab:La2}
scattering problem.
The latter authors in turn based their scattering problem
on the general AKNS problem given by Ablowitz, Kaup, Newell, and Segur
\cite{akns:74}.

It is well known that a single linear problem can give rise to a hierarchy of
integrable ODEs
\cite{Flasch:New}.
In this paper, we announce an extension of such hierarchies
to the discrete realm by deriving hierarchies of
integrable difference equations.

The plan of the paper is as follows. We recall the isomonodromy
problem for dP$\rm_I$ in section 2 for completeness and then show how
higher-order difference equations arise from it, giving examples
up to the sixth-order.

In section 3, we give continuum limits
of the second and fourth-order dP$\rm_I$ equations. In section 4,
we examine a linear problem associated with P$\rm_I$ and show that the
continuum limit found for the fourth-order dP$\rm_I$ is the next equation in
the P$\rm_I$ hierarchy.
We propose that our hierarchy of difference equations
is an integrable discrete version of such a hierarchy.

\section[2]{Construction of the dP$\rm_I$ hierarchy}

The linear problem associated with dP$\rm_I$ \cite{Josh:iso} is
\be
	\alpha_n w_{n+1}&=&\lambda w_{n} - w_{n-1},\label{lax1}\\
         \frac{\partial w_{n}}{\partial \lambda}&=&a_{n}w_{n+1}+b_{n}
w_{n},\label{lax2}
\ee
where $w, \alpha, a, b$ depend on a discrete variable $n$ and $w, a, b$
also depend on
the continuous variable $\lambda$ ($a,b$ rational in $\lambda$).

The aim of this section is to show that
dP$\rm_I$ is one of a whole
range
of higher order compatibility conditions arising from this
system.

The compatibility conditions of equations (\ref{lax1}) and (\ref{lax2}) are
\be
&&	b_{n+1}-b_{n-1} + \lambda
(\frac{a_{n+1}}{\alpha_{n+1}}-\frac{a_{n}}{\alpha_{n}})=0,\nonumber\\
&&	\frac{\lambda^2}{\alpha_{n}}
(\frac{a_{n+1}}{\alpha_{n+1}}-\frac{a_{n}}{\alpha_{n}})+\frac{\lambda}
{\alpha_{n}}
(b_{n+1}-b_{n})-\frac{1}{\alpha_{n}} (\alpha_{n}\frac{
a_{n+1}}{\alpha_{n+1}}-a_{n-1}+1)=0.\nonumber
\ee
Define
\be
	p_{n}=\frac{a_{n}}{\alpha_{n}}\hspace{1cm}\hbox{and}\hspace{1cm}
 q_{n}=b_{n}+b_{n-1}.\nonumber
\ee
We concentrate on results for $p_n$. Corresponding results for $q_n$ can be
obtained from the first compatibility condition above.

The two compatibility conditions collapse down into a single condition in
terms of $p_{n}$
\beqn
	\alpha_{n+1} p_{n+2} + (\alpha_{n}-\lambda^2) p_{n+1} +
(\lambda^2-\alpha_{n}) p_{n} - \alpha_{n-1} p_{n-1} + 2 = 0.\label{compcond}
\eeqn
Take $p_{n}$ to be a finite Laurent expansion in $\lambda$
\be
	p_{n}= \sum^{l}_{k=0} P_{k,n} \lambda^k,\nonumber
\ee
and substitute this expansion into (\ref{compcond}). (Wherever convenient
below we refer to $P_{k,n}$ as $P_k$.)
We then find
the following set of simultaneous equations for $P_{0,n},P_{1,n},...P_{l,n}$
\begin{eqnarray*}
        P_{l,n}-P_{l,n+1}&=&0,\\
        P_{l-1,n}-P_{l-1,n+1}&=&0,
\end{eqnarray*}
\[\alpha_{n+1} P_{k,n+2} + \alpha_{n} P_{k,n+1} - \alpha_{n} P_{k,n}
        -\alpha_{n-1} P_{k,n-1} + P_{k-2,n} - P_{k-2,n+1}=0\]
for $ 2 \leq k \leq l$, and
\begin{eqnarray*}
\alpha_{n+1} P_{1,n+2}+\alpha_{n} P_{1,n+1}- \alpha_{n} P_{1,n} -
        \alpha_{n-1} P_{1,n-1}&=&0\\
\alpha_{n+1} P_{0,n+2}+\alpha_{n} P_{0,n+1}-\alpha_{n}
        P_{0,n}-\alpha_{n-1} P_{0,n-1}+2&=&0
\end{eqnarray*}
for $k=0,1$.

Note that the equation for $P_{0}$ is inhomogeneous (i.e. it has a nonzero
forcing term) whereas the equation for $P_{1}$ is homogeneous. Also note that
the equations defining $P_{k}$ for even and odd $k$ may be
separated into otherwise identical equations.
These lead to two inconsistent
compatibility conditions for $\alpha$ unless one of these two subsequences of
$P_{k}$ vanishes identically. In fact, $P_l=0$ for odd $l$ is the
only possibility due to the inhomogeneity of the equation for
$P_{0}$.

We are therefore left with the following system
\begin{eqnarray*}
	P_{2m,n}-P_{2m,n+1}=0,
\end{eqnarray*}
\[\alpha_{n+1} P_{2k,n+2} + \alpha_{n} P_{2k,n+1} - \alpha_{n} P_{2k,n}
	-\alpha_{n-1} P_{2k,n-1}+ P_{2k-2,n}- P_{2k-2,n+1}
        =0\]
for $ 1 \leq k \leq m$, and
\begin{eqnarray*}
	\alpha_{n+1} P_{0,n+2}+\alpha_{n} P_{0,n+1}-\alpha_{n}
	P_{0,n}-\alpha_{n-1} P_{0,n-1}+2=0
\end{eqnarray*}
for $k=0$, where we have taken $l=2m$ ($m\in {\bf N}$).

Below we list the solutions for different
$m$ up to $m=3$.
The
$c_{i}$'s, $i=1,2,...,5$ are constants.

\begin{itemize}
\item $m=0$ yields the trivial linear nonautonomous equation
\be
	\alpha_{n} c_{2}=c_{1} +c_{0}(-1)^n - n\nonumber
\ee
with
\be
	P_{0,n}=c_2 \neq 0.\nonumber
\ee

\item $m=1$ yields the second-order equation
\be
	\alpha_{n} c_{3}(\alpha_{n-1}+\alpha_{n}+\alpha_{n+1})+\alpha_{n} c_{2}
	=c_{1} +c_{0}(-1)^n - n\label{dPII}
\ee
with
\be
	&&P_{0,n}=c_2+ c_3 (\alpha_n + \alpha_{n-1}),\nonumber\\
	&&P_{2,n}=c_3 \neq 0.\nonumber
\ee
In this case, we have recovered the more general version of dP$\rm_I$
\cite{Gram:Ram:Papa}.

\item $m=2$ yields the fourth-order equation
\be
	&&\alpha_{n} c_{4} (\alpha_{n+1}\alpha_{n+2}+\alpha_{n+1}^2 +2
	\alpha_{n}\alpha_{n+1}
	+ \alpha_{n-1}\alpha_{n-2}
        + \alpha_{n-1}^2+2\alpha_{n}\alpha_{n-1}\nonumber\\
	&&+\alpha_{n}^2+\alpha_{n-1}\alpha_{n+1})
	+\alpha_{n} c_{3}(\alpha_{n-1}+\alpha_{n}+\alpha_{n+1})+\alpha_{n}
	c_{2}\nonumber\\
	&&=c_{1} +c_{0}(-1)^n - n\label{4th}
\ee
with
\be
	&&P_{0,n}=c_2+ c_3 (\alpha_n + \alpha_{n-1})\nonumber\\
&&\hspace{1.2cm}+ c_4 (2 \alpha_n \alpha_{n-1}
+ \alpha_{n-1}^2 + \alpha_{n-1}\alpha_{n-2}+\alpha_n\alpha_{n+1}+\alpha_n^2),
\nonumber\\
	&&P_{2,n}=c_3+ c_4 (\alpha_n + \alpha_{n-1}),\nonumber\\
	&&P_{4,n}=c_4 \neq 0.\nonumber
\ee

\item $m=3$ yields the sixth-order equation
\be
	&&\alpha_{n} c_{5} (\alpha_{n+1} \alpha_{n+2} \alpha_{n+3}
	+2 \alpha_{n}\alpha_{n+1}\alpha_{n+2}
	+ \alpha_{n-1}\alpha_{n+1}\alpha_{n+2} + \alpha_{n+1}^3\nonumber\\
	&&+\alpha_{n+1}\alpha_{n+2}^2+2 \alpha_{n+2}\alpha_{n+1}^2
	+ 3 \alpha_{n+1}\alpha_{n}^2 + 3\alpha_{n}\alpha_{n+1}^2
        +\alpha_{n-1}\alpha_{n-2}\alpha_{n-3}\nonumber\\
	&&+2\alpha_{n}\alpha_{n-1}\alpha_{n-2}
	+\alpha_{n-2}\alpha_{n-1}\alpha_{n+1} + \alpha_{n-1}^3
	+\alpha_{n-2}^2\alpha_{n-1}+2\alpha_{n-1}^2\alpha_{n-2}\nonumber\\
	&&+3\alpha_{n-1}\alpha_{n}^2 + 3\alpha_{n-1}^2\alpha_{n}
	+4\alpha_{n-1}\alpha_{n}\alpha_{n+1} +\alpha_{n}^3
	+\alpha_{n-1}\alpha_{n+1}^2\nonumber\\
        &&+\alpha_{n-1}^2\alpha_{n+1})
	+\alpha_{n} c_{4} (\alpha_{n+1}\alpha_{n+2} + \alpha_{n+1}^2 +2
	\alpha_{n}\alpha_{n+1}
	+ \alpha_{n-1}\alpha_{n-2}\nonumber\\&&+ \alpha_{n-1}^2
+2\alpha_{n}\alpha_{n-1}+\alpha_{n}^2+\alpha_{n-1}\alpha_{n+1})
	+\alpha_{n} c_{3}(\alpha_{n-1}+\alpha_{n}+\alpha_{n+1})\nonumber\\
&&+\alpha_{n}
	c_{2}
	=c_{1} +c_{0}(-1)^n - n\nonumber
\ee
with
\be
	&&P_{0,n}=c_2+ c_3 (\alpha_n + \alpha_{n-1}) + c_4
(2 \alpha_n \alpha_{n-1}
+ \alpha_{n-1}^2 + \alpha_{n-1}\alpha_{n-2}\nonumber\\
&&\hspace{1.2cm}+\alpha_n\alpha_{n+1}+\alpha_n^2)
+c_5 (\alpha_n\alpha_{n+1}\alpha_{n+2}+2 \alpha_{n-1}\alpha_n\alpha_{n+1}+3
\alpha_n^2\alpha_{n-1}
\nonumber\\
&&\hspace{1.2cm}+ 3 \alpha_{n-1}^2\alpha_n
+\alpha_{n-2}^2\alpha_{n-1}+2 \alpha_{n+1}\alpha_n^2 +\alpha_n^3 +
\alpha_{n-1}^3+\alpha_n\alpha_{n+1}^2\nonumber\\
&&\hspace{1.2cm}+\alpha_{n-1}\alpha_{n-2}\alpha_{n-3} + 2
\alpha_n\alpha_{n-1}\alpha_{n-2} + 2 \alpha_{n-1}^2\alpha_{n-2}),\nonumber\\
	&&P_{2,n}=c_3 + c_4 (\alpha_n + \alpha_{n-1}) \nonumber\\
&&\hspace{1.2cm}+ c_5 (2 \alpha_n \alpha_{n-1}
+ \alpha_{n-1}^2 + \alpha_{n-1}\alpha_{n-2}+\alpha_n\alpha_{n+1}+\alpha_n^2),
\nonumber\\
	&&P_{4,n}=c_4+ c_5 (\alpha_n + \alpha_{n-1}),\nonumber\\
	&&P_{6,n}=c_5 \neq 0.\nonumber
\ee
\end{itemize}

This process of solving the system for increasing $m$ continues
indefinitely. At each stage, the order of the compatibility condition increases
by $2$.

\section[3]{Continuum limits}

In this section, we derive continuum limits of difference equations found in
the previous section.

\subsection{The case $m=1$}

Consider equation (\ref{dPII}).
If $c_{0}=0$, we recover the equation often referred to as dP$\rm_I$,
with P$\rm_I$ as one of its continuum limits.
If $c_{0}\neq 0$ however, the term $(-1)^n$ suggests an odd-even
dependence in $\alpha_n$.  This dependence must be taken into account to obtain
a meaningful continuum limit. This leads to the transformation
\be
	\alpha_{2k-1}=u_{k},\hspace{1cm}\alpha_{2k}=v_{k}.\nonumber
\ee
(This is similar to the limit pointed out by Grammaticos \textit{et al}
\cite{Gram:Nij:Papa:Ram:Sats} for a discrete version of the second
Painlev\'{e} equation generalised with an additional
$(-1)^n$ term.)

In our search for a continuum limit, we use the substitutions
\be
	u_{k} = 1 + h y(k h),\hspace{0.6cm}v_{k}=z(kh),
\hspace{0.6cm}t=k h,\hspace{0.6cm}c_3=-\frac{2}{r_1} h^{-3},\nonumber
\ee
and for ease of notation rename
\be
\mu=\frac{c_{1}+c_{0}}{c_{3}},\hspace{0.6cm}
\nu=\frac{c_{1}-c_{0}+1}{c_{3}},\hspace{0.6cm}
\sigma=-\frac{c_{2}}{c_{3}}\hspace{0.4cm}\hbox{and}\hspace{0.4cm}
\rho=-\frac{2}{c_{3}}.\nonumber
\ee
Then we find
\be
	z=\frac{\sigma+\nu-1}{2}
	-\frac{\nu +1}{2} y h
	+\frac{1}{2}(r_1 t+\nu y^2-\frac{\nu}{2}
	y_{t}-\frac{1}{2}y_{t})h^2\nonumber\\
	+\frac{1}{2}(\frac{r_1}{2} +\nu y y_{t}-r_1 t y-
	\nu y^3)h^3+O(h^4).\nonumber
\ee
Using the scalings
\be
\mu&=&-\frac{\sigma^2}{4}+\frac{\nu^2}{4}+\frac{\nu}{2}+\sigma-\frac{3}{4}
-r_2 h^3,\nonumber\\
\sigma&=&\frac{3}{2}+\frac{\nu^2}{2}+r_3 h^2,\nonumber\\
\nu&=&1+\frac{2}{3} r_4 h,\nonumber
\ee
in the limit $h\rightarrow0$ we are left with
\be
y_{tt}-2 y^3 + 2 r_4 y^2 -2 r_1 t y +2r_3 y+\frac{2}{3} r_1 r_4 t  +r_1 +2
r_2 =0.\nonumber
\ee
This is a scaled and translated version of the second Painlev\'{e} equation
(P$\rm_{II}$).

\subsection{The case $m=2$}

For simplicity, we restrict ourselves to the case $c_0=0$.
First, we illustrate the fact that second-order continuum limits are possible
even though difference equation (\ref{4th}) is fourth order.
Under the substitution
\be
\alpha_{n} = 1 + h^2  u(n h),\hspace{1cm}t=n h,\nonumber
\ee
and the scalings
\be
10+\frac{3 c_{3}}{c_{4}}-\frac{c_{1}}{c_{4}}+\frac{c_{2}}{c_{4}}=r_1 h^4,
\hspace{0.6cm}
20+\frac{3 c_{3}}{c_{4}}+\frac{c_{1}}{c_{4}}=r_2 h^2,\hspace{0.6cm}
\frac{1}{c_{4}}=r_3 h^5,\nonumber
\ee
equation (\ref{4th}) becomes
\be
(10+\frac{c_{3}}{c_{4}}) u_{tt} +(10-
\frac{c_{1}}{c_{4}})u^2+r_2 u +r_1+r_3 t=0\nonumber
\ee
in the limit $h\rightarrow0$.
This is a scaled and translated version of P$\rm_I$. However, this case
restricts the four degrees of freedom contained in the parameters of equation
(\ref{4th})
to three.

Scalings that maintain the full four degrees of freedom, i.e.
\be
&&10+\frac{3 c_{3}}{c_{4}}-\frac{c_{1}}{c_{4}}+\frac{c_{2}}{c_{4}}=r_1 h^6,
\hspace{0.6cm}
20+\frac{3 c_{3}}{c_{4}}+\frac{c_{1}}{c_{4}}=r_2 h^4,\nonumber\\
&&10 + \frac{c_{3}}{c_{4}}=r_3 h^2,\hspace{0.6cm}
\frac{1}{c_{4}}=r_4 h^7,\nonumber
\ee
lead, in the limit $h\rightarrow0$, to the fourth-order equation
\be
u_{tttt} +5 (u_{t})^2 + 10 u u_{tt} +r_3 u_{tt}+ 10 u^3 + 3r_3 u^2 +r_2 u +r_1
+r_4
t=0.\label{lim4thnoc}
\ee
We discuss the significance of this equation in the next section.

\section[4]{Discussion}
The aim of this section is to show that equation
(\ref{lim4thnoc}), the continuum limit for the case $m=2$, $c_0=0$, is the
next equation in the P$\rm_I$ hierarchy.

The continuum limit equations found for the case $m=1$ are P$\rm_I$ and
P$\rm_{II}$. These equations are considered integrable as they are
isomonodromy conditions for a
single-valued linear problem. Furthermore, they
may be viewed as equations lying
at the base of two integrable hierarchies of equations.
Consider the following isomonodromy problem associated with P$\rm_I$
\be
	\bf{v}_x=\pmatrix{0&1\cr
                   -u(x)-k^2&0\cr}\bf{v},\hspace{1cm}
	\bf{v}_k=\pmatrix{A&B\cr
                   C&-A\cr}\bf{v},\label{isoprob}
\ee
where $u$ is the potential, $k$ is the eigenvalue and
\be
A&=&-2 u_x k,\nonumber\\
B&=&(2r_2+4u)k-8k^3,\nonumber\\
C&=&-(2u_{xx}+2r_2u+4 u^2)k+(-2r_2+4u)k^3+8k^5.\nonumber
\ee
P$\rm_I$ is a compatibility condition for system (\ref{isoprob}) with these
choices of $A$, $B$, $C$. A hierarchy of compatibility conditions arises when
$A$, $B$ and $C$ are expanded to higher degree in $k$.
The next self-consistent expansions are
\be
A&=&-(\frac{s_0}{16}u_{xxx}+\frac{3
s_0}{8}uu_x+\frac{s_1}{4}u_x)k+\frac{s_0}{4}u_x k^3,\nonumber\\
B&=&(\frac{s_0}{8}u_{xx}+\frac{3 s_0}{8}u^2+\frac{s_1}{2}u-s_2)k-
(\frac{s_0}{2}u+s_1)k^3+s_0k^5,\label{exp5}\\
C&=&(\frac{s_0}{8}u
u_{xx}-\frac{s_0}{16}u_x^2+\frac{s_0}{4}u^3+\frac{s_1}{4}u^2-2x+s_3)k
\nonumber\\&&
+(\frac{s_0}{8}u_{xx}+\frac{s_0}{8}u^2+\frac{s_1}{2}u+s_2)k^3
+(-\frac{s_0}{2}u+s_1)k^5-s_0k^7,\nonumber
\ee
where $s_0, s_1, s_2, s_3$ are constants. A straight forward calculation shows
that equation
(\ref{lim4thnoc}) is a compatibility condition of
(\ref{isoprob}) with $A$, $B$, $C$ expanded as in (\ref{exp5})
(renaming
$s_0=-\frac{32}{r_4}$, $s_1=-\frac{8r_3}{r_4}$, $s_2=\frac{2r_2}{r_4}$,
$s_3=-\frac{2r_1}{r_4}$).

\section[]{Acknowledgements}
The research reported in this paper was made possible through an Australian
Postgraduate Award and
 supported by the Australian Research
Council.


\begin{thebibliography}{99}

\bibitem{pp:acta}
P.~Painlev\'e.
\newblock Sur les \'equations diff\'erentielles du second ordre et d'ordre
  sup\'erieur dont l'int\'egrale g\'en\'erale est uniforme.
\newblock {\em Acta Math.}, 25:1--85, 1902.

\bibitem{gambier:acta}
B.~Gambier.
\newblock Sur les {\'equations} {diff\'erentielles} du second ordre et du
  premier {degr\'e} dont {l'int\'egral} {g\'en\'eral} est {\`a} points
  critiques fixes.
\newblock {\em Acta Math.}, 33:1--55, 1910.

\bibitem{fuchs:annalen}
R.~Fuchs.
\newblock {\"Uber} lineare homogene differentialgleichungen zweiter or ordnung
  mit drei im endlich gelegene wesentlich {singul\"aren} stellen.
\newblock {\em Math. Annalen}, 63:301--321, 1907.

\bibitem{okamoto:86}
K.~Okamoto.
\newblock Isomonodromic deformation and {Painlev\'e} equations, and the
  {Garnier} system.
\newblock {\em J. Fac. Sci. Univ. Tokyo Sec. IA Math}, 33:575--618, 1986.

\bibitem{Gram:Ram:Papa}
B.~Grammaticos, A.~Ramani, and V.~Papageorgiou.
\newblock Do integrable mappings have the {Painlev\'e} property?
\newblock {\em Phys. Rev. Lett.}, 67:1825--1828, 1991.

\bibitem{Ram:Gram:Hiet}
A.~Ramani, B.~Grammaticos, and J.~Hietarinta.
\newblock Discrete versions of the {Painlev\'e} equations.
\newblock {\em Phys. Rev. Lett.}, 67:1829--1832, 1991.

\bibitem{Josh:iso}
N.~Joshi, D.~Burtonclay, and R.~Halburd.
\newblock Nonlinear nonautonomous discrete dynamical systems from a general
  discrete isomonodromy problem.
\newblock {\em Lett. Math. Phys.}, 26:123--131, 1992.

\bibitem{Ab:La}
M.J. Ablowitz and J.F. Ladik.
\newblock Nonlinear differential-difference equations.
\newblock {\em J. Math. Phys.}, 16:598--603, 1975.

\bibitem{Ab:La2}
M.J. Ablowitz and J.F. Ladik.
\newblock Nonlinear differential-difference equations and fourier analysis.
\newblock {\em J. Math. Phys.}, 17:1011--1018, 1976.

\bibitem{akns:74}
M.J. Ablowitz, D.J. Kaup, A.C. Newell, and H.~Segur.
\newblock The inverse scattering transform --- {Fourier} analysis for nonlinear
  problems.
\newblock {\em Stud. Appl. Math}, 53:249--315, 1974.

\bibitem{Flasch:New}
H.~Flaschka and A.C. Newell.
\newblock Monodromy- and spectrum-preserving deformations {I}.
\newblock {\em Commun. Math. Phys.}, 76:65--116, 1980.

\bibitem{Gram:Nij:Papa:Ram:Sats}
B.~Grammaticos, F.W. Nijhoff, V.~Papageorgiou, A.~Ramani, and J.~Satsuma.
\newblock Linearization and solutions of the discrete {Painlev\'e} {III}
  equation.
\newblock {\em Phys. Lett. A}, 185:446--452, 1994.

\end{thebibliography}
\end{document}